\documentclass{aa}

\usepackage[varg]{txfonts}
\usepackage{multirow}
\usepackage{natbib}
\bibpunct{(}{)}{;}{a}{}{,} 
\usepackage{xcolor}
\usepackage{breqn}
\usepackage[colorlinks=true]{hyperref}
\usepackage[xindy]{glossaries}
\hypersetup{linkcolor=red,citecolor=blue,filecolor=cyan,urlcolor=green}
   
\usepackage{graphicx}
\usepackage{verbatim}

\newcommand*\subtxt[1]{_{\mathrm{#1}}}
\DeclareRobustCommand\_{\ifmmode\expandafter\subtxt\else\textunderscore\fi}
\begin{document}

   \title{The impact of Post-Newtonian effects on massive black hole binary evolution at $\sim 1000\,R_{sch}$ separations}

 \author{Branislav Avramov\inst{1}, 
           Peter Berczik \inst{2, 3}, 
          \and
          Andreas Just\inst{1}
          }

   \institute{Zentrum f\"ur Astronomie der Universit\"at Heidelberg, Astronomisches Rechen-Institut,  M\"onchhofstr. 12-14, 69120 Heidelberg, Germany  \\
         \email{branislav.m.avramov@gmail.com}
         \and    
         National Astronomical Observatories and Key Laboratory of Computational Astrophysics, Chinese Academy of Sciences, 
         20A Datun Rd., Chaoyang District, Beijing 100101, China
         \and  
         Main Astronomical Observatory, National Academy of Sciences of Ukraine, 27 Akademika Zabolotnoho St., 03143 Kyiv, Ukraine 
        }

   \date{Received ??; ??}

 
  \abstract
   {}
   {We study the impact of Post-Newtonian correction terms on the energetic interaction between a gravitational wave (GW)-emitting supermassive black hole (SMBH) binary system and incoming stars via three-body scattering experiments. }
   {We use the AR-chain code to simulate with high accuracy the interactions between stars and an SMBH binary at separations of $\sim 1000\,R_{sch}$. For all of the interactions, we investigate in detail the energy balance of the three-body systems, using both Newtonian and Post-Newtonian expressions for the SMBH binary orbital energy, taking into account the GW emission by the binary. }
   {We find that at these separations, purely Newtonian treatment of the binary orbital energy  is insufficient to properly account for the SMBH binary orbital evolution. Instead, along with GW emission, even terms in the PN-corrections must be included in order to describe the energy change of the binary during the stellar interaction.  }
   {}

\titlerunning{Impact of Post-Newtonian effects on massive black hole binary evolution}
\authorrunning{B. Avramov et al.}
\maketitle

%

\section{Introduction}

Supermassive black hole (hereafter SMBH) mergers are promising candidates for gravitational wave (GW) detections with the LISA mission \citep{LISA2017} and the pulsar timing arrays \citep{Reardon2016, Desvignes2016, Arzoumanian2016, Verbiest2016}. It is expected that following galactic mergers, the SMBHs which reside in the center of each galaxy form binary systems which would eventually coalesce, emitting a burst of GWs. \citet{Begelman1980} famously described the three main phases of an SMBH merger. In the first phase, dynamical friction drives the evolution of the binary, bringing it close enough to form a binary system. In the second phase, known as the hardening phase,  the SMBH separation is at parsec scales. At these separations, dynamical friction is no longer effective and instead individual three-body interactions of stars with the SMBH binary drive its evolution in a process known as stellar hardening \citep{Yu2002}. Finally, when the binary separation is at milliparsec scales, GW emission effectively carries away energy and angular momentum from the binary, bringing it to coalescence.

  During the hardening phase of a SMBH binary merger,  the main mechanism of energy loss for the binary is via energetic stellar interactions, during which incoming stars on radial orbits are able to extract energy from the binary with the gravitational slingshot effect. In order for the energy extracted from the binary to be significant enough to measurably contribute to the hardening, the star needs to experience very close interactions with the binary (with the closest approach distance on the order of the binary semi-major axis, $r_p \lesssim 3a_{bh}$, \citet{Yu2002}).  The proper treatment of this physical process in simulations is of crucial importance since this is the primary method of bringing the black holes close enough in order to reach coalescence. However,  it may be problematic and cumbersome to include in galactic-scale simulations, due to the very high spatial resolution which is necessary to properly resolve all of the close interactions (compared to galactic scales which are orders of magnitude larger). Therefore, treatment of these interactions with direct integration might be out of reach for studies that do not have access to large and efficient computational resources. Instead, they may use semi-analytical estimates of hardening rates which can then be used to model the evolution of SMBH binaries without actually resolving the encounters themselves \citep[see e.g.][]{Sesana2015}. 

\par On the other hand, direct N-body codes are able to calculate forces during the encounters up to a high degree of accuracy, but are often unable to accurately analyze the individual encounters between the stellar and BH particles, due to resolution constraints and exceptionally high computational cost. Therefore, they are usually unable to give precise estimates of the specific energy changes in a single encounter, instead focusing on cumulative effects from many interactions.

\par Instead, 3-body scattering experiments provide a computationally efficient way to study the nature of these interactions in a multitude of different configurations and scenarios and to investigate these effects for different masses of the binary members and orbital configurations  \citep[e.g.,][]{Bonetti2020,Rasskazov2019,Sesana2008, Sesana2006, Quinlan1996}. They can be especially useful as a precursor to N-body runs, as a way to gain insight into the cumulative, as well as average energy changes from stellar encounters, without the presence of other physical effects from the galactic environment which may affect the result, such as gravitational torques from overdensities in the distribution, two-body relaxation effects and the gravitational impact of the system itself.

\par In numerical simulations, the GW emission of the binary, as well as other relativistic effects are reproduced by adding Post-Newtonian (PN) corrections up to a certain order to the binary equations of motion \citep[see][for reviews on the topic]{Blanchet2014, Will2011}. While this approach is standard practice, it is still not fully understood if the introduction of Post-Newtonian terms in the equations of motion will affect the star-binary energy exchange, and if so, if the effect would be measurable. This is especially interesting to consider during the phase of the merger when gravitational wave emission becomes comparable to the binding energy gained from stellar interactions. During this transitional period, both stellar hardening and PN terms induce secular, as well as periodic changes in the orbital motion of the binary and the interplay between these two effects in N-body studies has not yet been thoroughly explored in the literature. 
\par Therefore, in this work we present the results of a number of three-body scattering experiments, performed using the highly accurate AR-chain integration method \citep{Mikkola2006, Mikkola2008}. The goal of these experiments was to provide valuable insight into the energy exchange at a time when the PN effects are non-negligible, enabling an overview of dominant terms in the energy balance when the binary is at  $\sim 1000\,R_{sch}$ separations ($R_{sch}$ is the Schwarzschild radius of the combined mass $m=m\_1 + m\_2$ of the SMBH binary).

\par The text is structured as follows. In Section \ref{sec:num} we describe the simulation parameters and describe the inital setup of the code. Section \ref{sec:results} contains our results. Specifically, in Section \ref{sec:merger} we briefly  describe the merger rate detected in the runs. In Section \ref{sec:energy} we present our main findings and discuss the relevance of PN terms in this phase of the merger. In Section \ref{sec:conc} we summarize and discuss our findings. Finally, in Appendix \ref{sec:6_vini}, we present the derivation of the condition  employed for the initial velocity of the stellar particles.

\section{Numerical description, initial conditions and setup}
\label{sec:num}
\par The three-body simulations were performed with the regularized AR-chain code \citep{Mikkola2006, Mikkola2008}, featuring two SMBH particles and a single stellar particle. The code includes PN corrections to the  equations of motion up to order 2.5PN, in order to account for GW emission. 
Each simulation run consists of three particles, two SMBH particles situated near the centre of mass of the system and a stellar particle, starting as a distant third body. Since the interactions between the particles are regularized to avoid singularities when $R \rightarrow 0$, we used no gravitational softening between the star and the black holes, as well as between the black holes themselves, allowing for arbitrarily close approaches. 

The masses, positions and velocities of the black holes were obtained using the system from \citet{Khan2016} at different times of the original simulation. In that work,  a massive galaxy merger at redshift $z\sim 3.5$ was identified and followed using the Argo cosmological simulation \citep{FeldmannMayer2015, Fiacconi_etal2015}. 
At a time which we refer to as the initial time $t\_{ini}$, a static particle-splitting procedure was performed in order to increase the particle number, and two SMBH particles with masses  $m\_{1} = 3 \times 10^{8} M_{\odot}$ and $m\_{2} = 8 \times 10^{7} M_{\odot}$ were introduced at the local minima of the gravitational potential of the galactic cores. 
    The system was evolved further using the GASOLINE code \citep{Wadsley2004} and during the final stages of the merger, the galaxy merger remnant had a gas fraction of only 5\%. At time $t\_{PN} = t\_{ini}+21.5$ $\mathrm{Myr,}$ the remaining gas particles were turned into star particles,  a spherical region of 5 kpc around the most massive SMBH was extracted, the softening was further reduced, and the PN terms were turned on. At this stage, the separation between the black holes was $\sim 300$ $\mathrm{pc}$.   This system was then further evolved using the direct N-body code $\varphi$-GPU \citep{Berczik2011}. Integration was continued until the merger of the SMBH particles was induced by the PN corrections (up to order 3.5) in the equations of motion. For more details on the simulation setup, we refer the reader to \citet{Khan2016}.  For more details on the properties of the system, we refer the reader to \citet{Avramov2021}. 
    


\par For the scattering simulations we used the same N-body units as in the original simulation (see Table \ref{tab:ini_cond} for the parameters). We performed two different sets of scattering experiments, corresponding to different times of the original N-body run. The first set of runs  is initialized when the binary semi-major axis is $a_0 = 4.6\times 10^{-2} \, \mathrm{pc}$, corresponding to a separation of $1277 \, R_{sch}$.  This separation corresponds to time $t_1 = t\_{PN} + 7.75 \, \mathrm{Myr} $ of the original run. At this point in time, we expect the gravitational wave emission to be measurable, but not dominant with respect to stellar hardening. The second set of runs is initialized when the binary semi-major axis is $a_0 = 3.15\times 10^{-2} \, \mathrm{pc}$, corresponding to a separation of $874 \, R_{sch}$ and  time  $t_2 = t\_{PN}+ 9.3 \, \mathrm{Myr} $ in the original run. At this point in time we expect gravitational wave emission to be dominant with respect to stellar hardening.  
The stellar particle is assigned a mass several orders of magnitude smaller than the mass of the less massive black hole, and is positioned at a large initial distance from the binary system $D_0= 100 \, \mathrm{pc}$. Every stellar particle is assigned the same mass value of  $91,800 \, \mathrm{M_{\odot}}$.

\par A simulation run is ended if the stellar particle becomes unbound. A stellar particle is considered unbound if its total energy becomes positive, $E_*>0$, and if it reaches a distance $R>2D_0$ from the center of mass. Both of these conditions need to be met in order to stop the run. Otherwise, a run is stopped if the star does not experience large energy changes over a sufficiently large timescale at $t\_{max}$, or if there is a merger event between the star particle and the black hole binary.  After each encounter, the system is reset, the binary is given the same initial conditions and orbital parameters as before, and the next stellar particle is generated. 

\par The initial conditions for stellar particles are calculated  in such a way so that the stellar particles are distributed spherically-symmetric around the SMBH binary.   Using standard random number generation, stars are distributed evenly in space on a spherical shell surrounding the binary with a radius $R = D_0$. 
For the initial velocities, a unit vector is drawn from an isotropic distribution, which determines the angle $\alpha$ between the radius vector and the velocity vector of the stellar particle.
In order to make sure that our encounters experience strong interactions with the black hole binary, we need to carefully choose the magnitude of the initial velocity of the stellar particles.  We are interested in particles which have closest approaches comparable to the initial semi-major axis $a_0$ of the binary:
\begin{equation}
\label{eq:per_cond}
r\_p<2a_0,
\end{equation}
where $r\_p$ is the pericenter distance in the two-body Keplerian approximation of the star-SMBH binary system.  
From this condition, we can obtain an equivalent condition for the velocity amplitude $V_{i}$ of the stellar particle in terms of the escape velocity of the star $V\_{esc} = \sqrt{2m/D_0}$  (here and hereafter the gravitational constant is set to unity, $G=1$, for simplicity) given by   
\begin{equation}
\label{eq:v_cond}
 \left(\frac{V_{i}}{V\_{esc}}\right)^2 \leq \frac{2a_{0}(D_{0}-2a_{0})}{D_{0}^2\sin^2\alpha-4a_0^2}
 \quad\mbox{for}\quad |\sin\alpha|>\frac{2a_{0}}{D_{0}}.
\end{equation}
The full derivation of this condition is given in Appendix \ref{sec:6_vini}.
The value of $V_{i}$ is chosen from a flat distribution in the allowed range.




 The first set of runs, corresponding to initial time $t_1$, consisted of $1000$ scattering experiments with output generated at a timestep of $\Delta t=1550 \, \mathrm{yr}$. The second set of runs consisted of  2000 encounters in total, and the output frequency was lowered since snapshots were generated  at the timestep of $\Delta t=15500 \, \mathrm{yr}$. The sample size for the second set of runs was increased by a factor of two  in order to investigate if the results are affected by the low number statistics. It is important to note  that the output frequency for both sets of runs is several orders of magnitude lower than the output frequency needed to actually resolve the orbit of the binary, which has an initial period $P\_0$ of 50\,yr and 27\,yr, respectively (see Table \ref{tab:ini_cond}).  This means that while the time resolution of the output data is significantly increased with respect to the original study, it is still insufficient to resolve the energetic encounter itself during post-processing.

\begin{table*}\centering
\caption{In this table we present the initial conditions and parameters of the runs.  }

{\begin{tabular}{|c|c|c|c|c|c|c|c|c|c|c|c|}
\hline 
$N$ & $\Delta t$ &  $t\_{0}$ &  $t\_{max}$ &   $m\_{1}$ & $m_{2}$ & $m\_{*}$ & $D\_0$ & $a\_0$ & $e\_0$ & $P\_0$ & PN terms   \\
\hline
1000 & $10^{-3}$ & 5 & 10 & $3.3\times10^{-3}$ & $8.7\times10^{-4}$ & $ 10^{-6}$ & 0.1 & $4.6 \times 10^{-5}$ & $0.13$ & $3.21\times10^{-5}$ & 1+2+2.5    \\
\hline
2000 & $10^{-2}$ & 6 & 10 & $3.3\times10^{-3}$ & $8.7\times10^{-4}$ & $ 10^{-6}$ & 0.1 & $3.15 \times 10^{-5}$ & $0.15$ & $1.73\times10^{-5}$ & 1+2+2.5   \\
\hline
\end{tabular}}
\tablefoot{All values are in N-body units ($G=1$, M.U. = $9.18 \times 10^{11} \, \mathrm{M_{\odot}}$ and L.U. = $1 \, \mathrm{kpc}$ leading to T.U. = $1.55 \, \mathrm{Myr}$ and  V.U. = $628.3 \, \mathrm{km\, s^{-1}}$). The columns represent in left-to-right order: number of runs per simulation set,  output frequency of run, initial time of run expressed in the original data timing scheme, maximum possible time of run, major black hole mass, minor black hole mass, stellar mass, initial stellar distance from the binary center of mass, initial semi-major axis of binary, initial orbital eccentricity of binary, initial orbital period of binary, order of PN terms included. }
\label{tab:ini_cond}
\end{table*}




\section{Results}
\label{sec:results}
\subsection{Merger rate}
\label{sec:merger}
\par Previously described parameter values are set up in such a way that is expected to be ideal for highly effective energy exchange between the stellar and SMBH particles. This is in large part due to the fact that the initial velocity condition given by Eq. \ref{eq:v_cond} guarantees close approaches to the binary. Normally, the value of the  softening determines the closest possible approach between the particles. Since the  softening in the AR-chain is set to zero for all particles, this artificial limit is not present and allows for arbitrarily close interactions between all particles.  However, this can also lead to merger events, which should be excluded from any energy exchange analysis.  An interaction between a stellar and black hole particle is registered as a merger when their separation becomes smaller than four times the Schwarzschild radius of the black hole involved in the interaction. The merger rates obtained from the simulations for the cases of  $t_1$ and $t_2$ are $2.01\%$ and $2\%$, respectively. These events are therefore excluded from further analysis, since in reality they would result in direct plunges of stars into one of the black holes.

\subsection{Energy balance}
\label{sec:energy}
\par
In this section, we will analyze in detail the energy balance of the system by examining  the different energy terms which constitute the total energy of the three-body system for each run and investigate the impact of PN terms to the results.

\par The equations of motion of the two black hole particles have been corrected for Post-Newtonian effects up to order 2.5PN, according to the formulas found in \citet{Blanchet2003}. However, it is important to note that the treatment of stellar particles remains strictly Newtonian. The only dissipative term in the equations of motion of the black holes is the 2.5PN term, corresponding to the GW radiation-reaction term \citep{Peters1963, Peters1964}, while the other PN terms (namely 1PN and 2PN) are conservative and maintain the conservation of energy of the SMBH binary.  The simulation employs black holes without spin, so the 1.5PN term is equal to zero. Because of this,  we expect the 2.5PN term to be the only one to effectively take away net energy from the system. 
\par 
The relative energy error of our simulation runs is in the range of  $10^{-13}$--$10^{-9}$,  several orders of magnitude smaller than the values of the energy components of the system. Therefore, we expect that when corrected for the 2.5PN term, the total energy of the simulated 3-body system is conserved to a high degree, and the energy exchange is well resolved.  Let us now provide an overview of all the energy components.  We will assume the binary center of mass is at rest at the origin  and treat the binary as a single massive body when computing the potential exerted by the binary on the star. These are justified approximations, since the initial ($D_0$) and final distance of the star ($ > 2D_0$) is much larger than the binary orbital separation, and the mass of the stellar particle is more than 4000 times smaller than the mass of the SMBH binary.

Without taking into account non-dissipative PN terms, the energy change (final -- initial values) of the black holes is given by

\begin{equation}
-\Delta E\_{Newt}  = \Delta E\_{*} + E\_{GW}, 
\label{eq:e_balance}
\end{equation}

where $E\_{BH}=E\_{Newt} $ is the total Newtonian energy of the binary defined as the sum of its potential and kinetic energy, $E\_{*}$ is the total energy of the star and $E\_{GW}=\Delta E\_{2.5PN}$ is the energy emitted by gravitational radiation during the simulation. Equation \ref{eq:e_balance} tells us that we expect that the binding energy change of the binary corresponds to  a high degree of accuracy to the sum of the energy change of the star particle and the energy emitted by gravitational radiation.   The instantaneous energy emitted via gravitational waves $E_{GW}$ corresponds to the time derivative of the radiation-reaction term and is calculated using the energy balance correction in the quadrupole approximation \citep{Blanchet2003}

\begin{equation}
E\_{2.5PN}= \frac{8m^3\dot{r}(\mu/m)^2}{5c^5r^2}v^2, \\
\end{equation}

where $\mu$ is the reduced binary mass $\mu = (m_1m_2)/(m_1+m_2)$, $r$ and $v$ are the relative distance and velocity of the SMBHs. The corresponding GW emission is then cross-checked with the orbit-averaged formulas from \citet{Peters1963, Peters1964}.

\par The left panels of Fig. \ref{fig:bind_corr} show the energy balance of all encounters according to  Eq. \ref{eq:e_balance} for the two sets of simulations. The black line corresponds to an ideal situation with no missing energy, when the energy change of the binary is exactly equal to the stellar energy changes. The color coding corresponds to Newtonian orbital eccentricity changes of the SMBH binary, calculated using the  Runge-Lenz vector. If Eq. \ref{eq:e_balance} can be used, all points should fall onto the one-to-one black line.  However, the plots show  that there is an energy discrepancy for each individual run, as shown by the vertical distance of the points from the black line.   The color coding shows that the value of the energy discrepancy is correlated to the eccentricity change of the binary. Positive changes in eccentricity lead mostly to a negative discrepancy in the energy balance, and vice-versa. 

\par The same energy discrepancy is noticeable when looking at the cumulatively summed terms of all of the interactions. An overview of the cumulatively summed different energy terms for both sets of simulation runs can be seen in Figure \ref{fig:cum_spec_corr} (left panels), as a function of specific energy change of the stars. If Eq. \ref{eq:e_balance} is satisfied, the blue and dashed magenta line, as well as the red and  green, dashed lines should match to a very high degree of accuracy. Then, we would expect that the energy difference between all terms (gray dash-dotted line) remains at zero at all energies.  However, we notice from the figure that this is not the case. In both simulation sets, there is a non-negligible energy discrepancy denoted by the gray dash-dotted line, which represents energy not accounted for by the energy balance equation. In the case of $t_1$, the cumulative sum of the energy discrepancies is positive, while when $t_2$ the summed energy discrepancies are negative.  This energy discrepancy is at the same order of magnitude as the other energy terms and therefore not negligible. As previously noted, the energy error of the simulation is a few orders of magnitude smaller than any value of the energy terms, and therefore cannot be responsible for the energy  discrepancy. 

\subsection{Post-Newtonian energy correction}

\par The presence of the energy discrepancy points to the fact that strictly Newtonian definitions of the binary orbital energy may be insufficient to account for the energy evolution of the system.   As previously described, the even PN terms are non-dissipative and we did not previously include them in the equation denoting the energy changes (Eq. \ref{eq:e_balance}), since they do not carry away energy from the system. Instead, the 1PN and 2PN terms are connected to correction terms for the energy of the binary. For eccentric orbits they depend on the orbital phase of the binary, which leads to a dominating oscillatory contribution when calculating energy differences (before and after the interaction). The long-term drift due to these terms is much smaller, leading to corrections of the order of $\sim R\_{sch}/r$.

\par 
The leading PN correction term to the binary orbital energy is given by  \citep{Blanchet2003}: 
\begin{align}
\label{eq:epn1}
E\_{1PN} = \frac{\mu}{c^2}&\left\{\frac{3v^4}{8}-\frac{9m_1m_2v^4}{8m^2}+ \frac{m^2}{2r^2}\right. \\
+&\left.\frac{m}{r}\left(\frac{m_1m_2\left(\vec{n}\cdot\vec{v}\right)^2}{2m^2} +\frac{3v^2}{2}+\frac{m_1m_2v^2}{2m^2} \right) \right\} \nonumber, 
\end{align}
where $\vec{n}$ is the normalized radius vector $\vec{n} = \vec{r}/r$, and $\vec{n}\cdot\vec{v}$ denotes the standard scalar product between the normalized radius and velocity vector. This leads to the total energy of the binary
\begin{equation}
\label{eq:e_bin_pn}
E\_{BH} = E\_{Newt} +E\_{1PN}.
\end{equation}
\par
\begin{figure*}
    \centering
\includegraphics[width=1\textwidth]{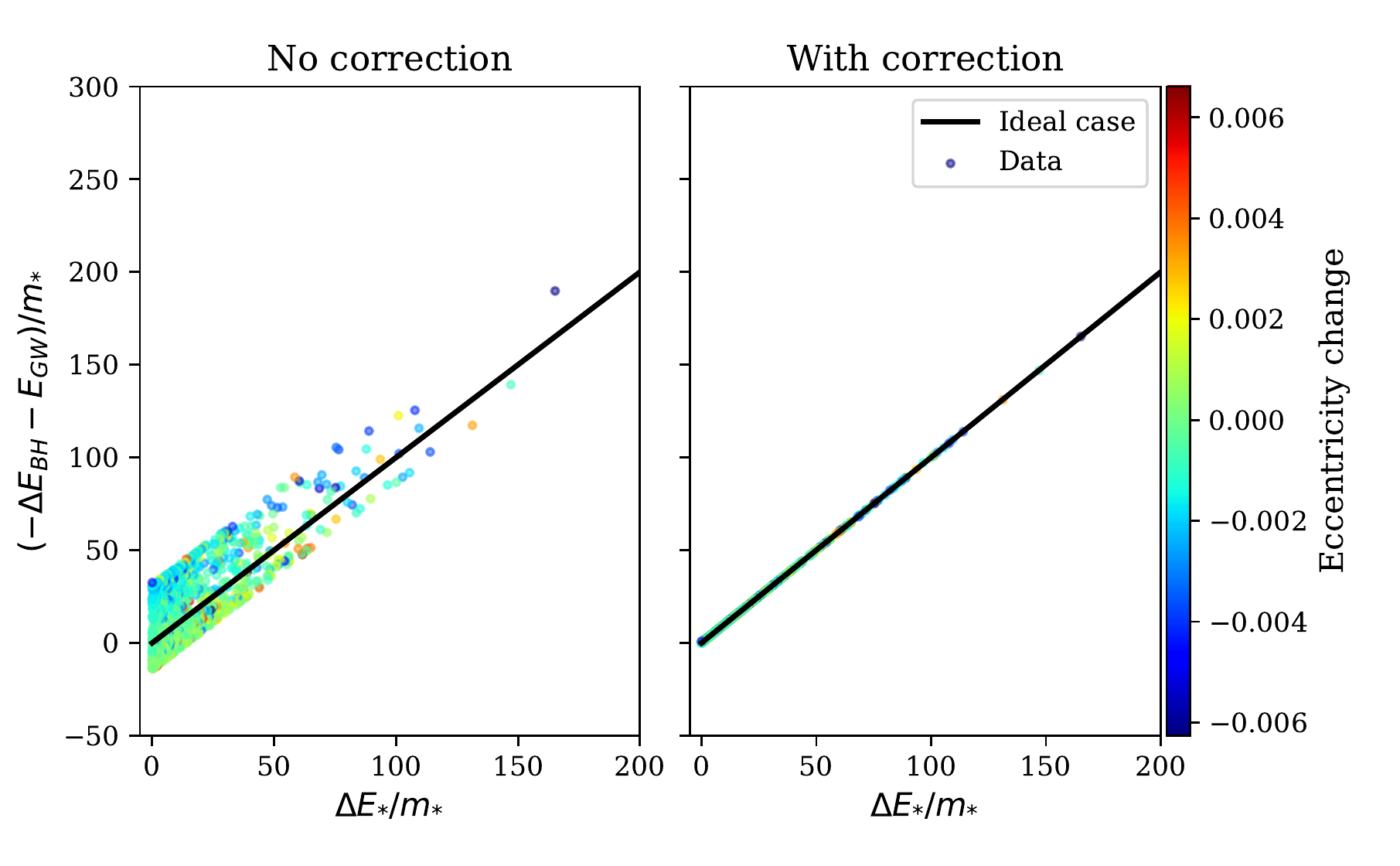} 
\includegraphics[width=1\textwidth]{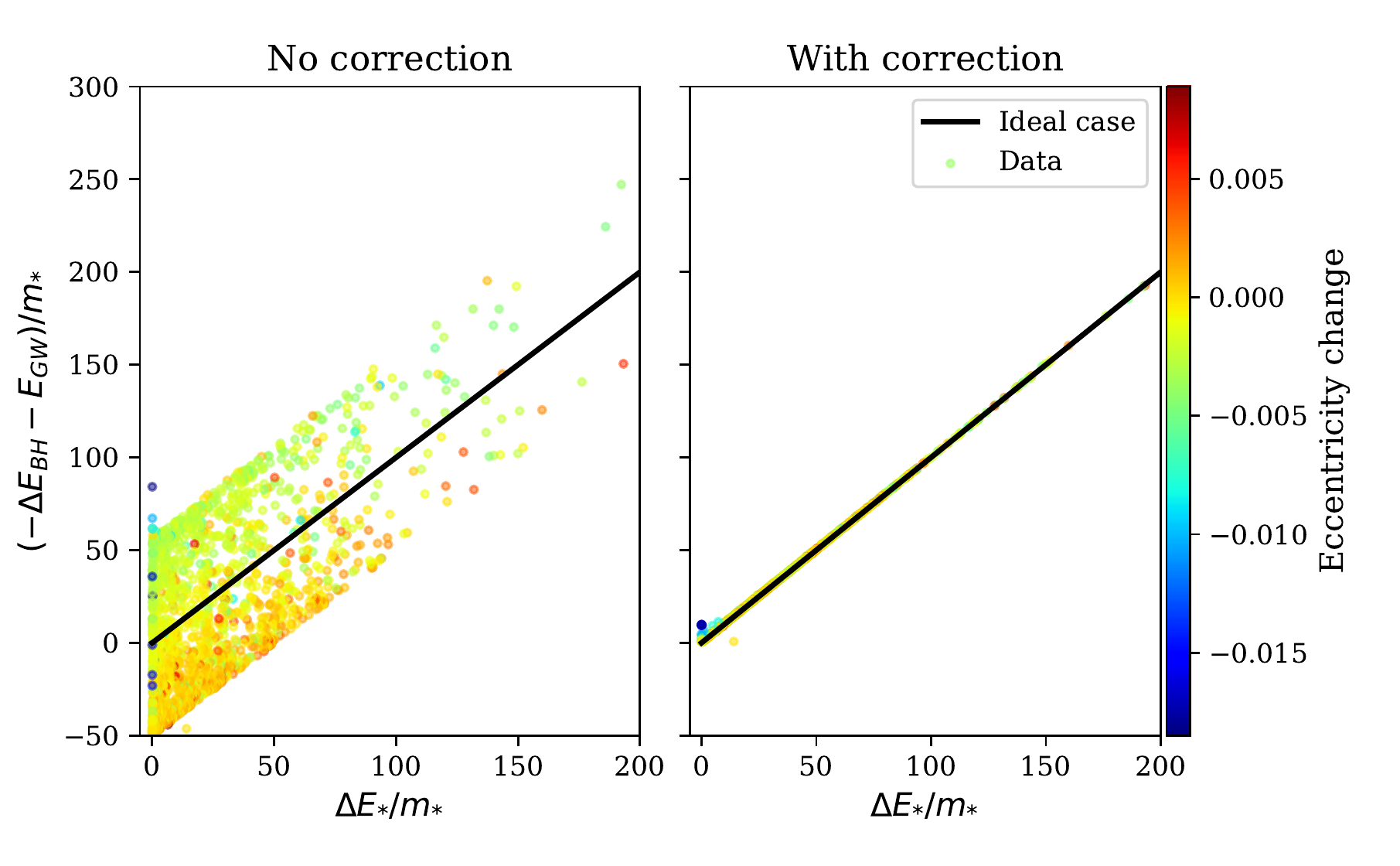}
\caption{SMBH binary orbital energy change normalized to the stellar mass, corrected for GW emission, as a function of specific energy change of the star, for initial time $t_1$ (top panels) and $t_2$ (bottom panels), without (left) and with (right) the $E_{1PN}$ correction term. Each point corresponds to one scattering experiment. The black line shows the ideal case when the quantity on the y-axis is identical to the quantity on the x-axis, corresponding to no additional energy corrections.} 
\label{fig:bind_corr}
\end{figure*}

 If the equations of motion are corrected up to the 2PN term, the energy of the binary $E\_{BH}$ is fully conserved. However, since the equations of motion also include the dissipative 2.5PN term, this energy is conserved only in the sense that its time derivative computed through the fully PN-corrected equations of motion is exactly equal to the effect of the 2.5PN term \citep{Blanchet2003}. This leads to the final expression for stellar energy change: 
\begin{equation}
\label{eq:e_star}
\Delta E\_* = -\Delta E\_{Newt} - \Delta E\_{1PN} -E\_{GW}.
\end{equation}
On the right side of  Fig. \ref{fig:bind_corr} we present again the energy changes after the run, but using the updated energy balance equation Eq. \ref{eq:e_star}. We immediately notice  that the energy discrepancies seen in the left side of the plot come from omitting the $E_{1PN}$ correction term, and after the correction we obtain much better energy conservation, with all points falling almost exactly on the black line.  At the later time, there are still small discrepancies for some encounters visible, which would disappear, when the 2PN corrections are also taken into account.

\begin{figure*}
\includegraphics[width=\textwidth]{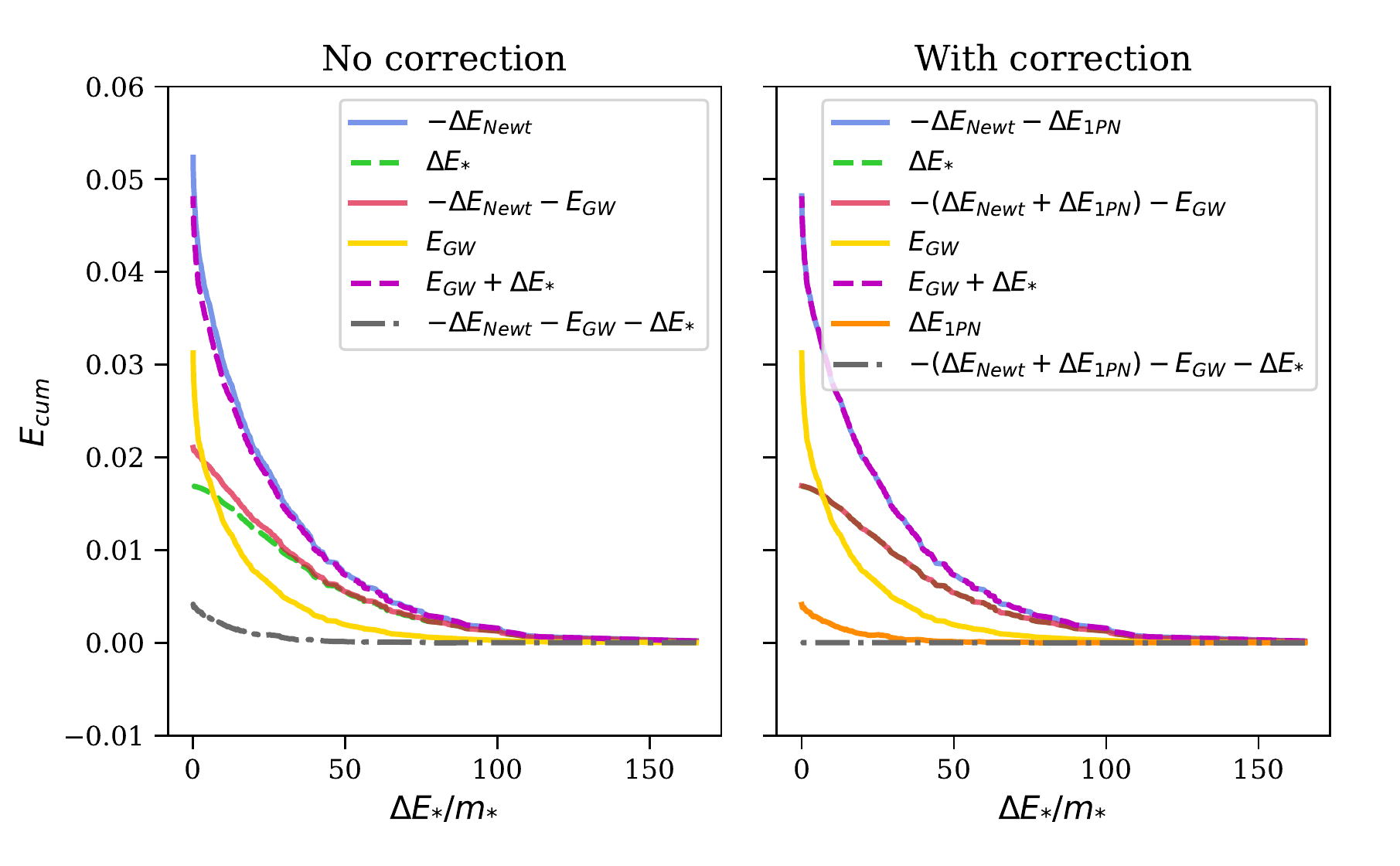}
\includegraphics[width=\textwidth]{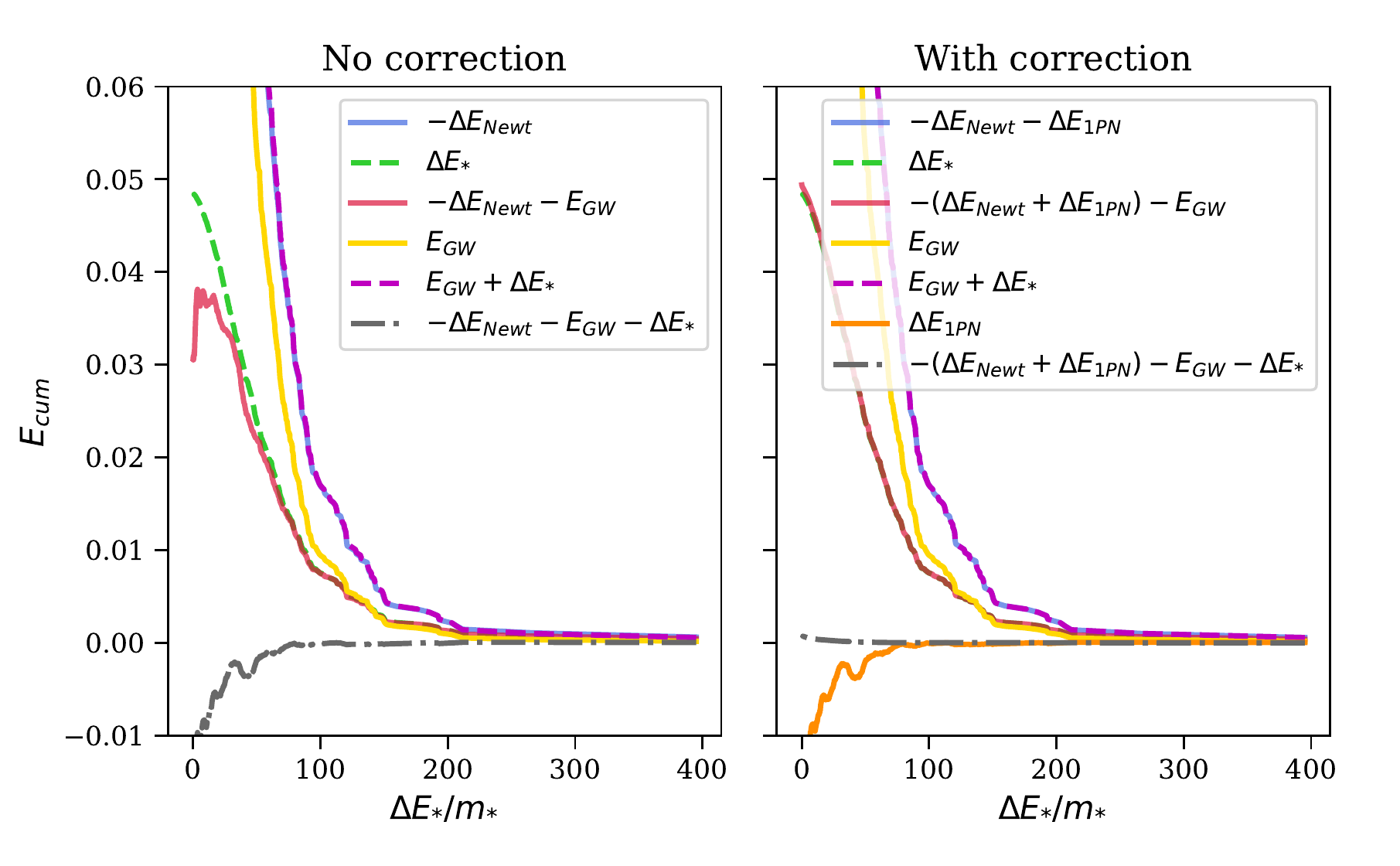}
\caption{Cumulative energy terms as a function of specific energy change, for the simulation set at initial time $t_1$ (top panels) and $t_2$ (bottom panels), without (left) and with (right) the $E_{1PN}$ correction term. 
The deviation of the grey line from zero in the bottom right plot corresponds to the 2PN energy correction, which is still much smaller than the 1PN correction.}
\label{fig:cum_spec_corr}
\end{figure*}

On the right side of Fig. \ref{fig:cum_spec_corr} we can see the updated cumulative energy term plots which include the correction for the leading order 1PN energy term to the energy balance. The first order correction is clearly sufficient to successfully resolve the energy discrepancy for $t_1$ when the separation between the black holes is $\approx 1300 \, R_{sch}$. This is shown by the gray dash-dotted line which is now consistently at zero for all encounters, proving that Eq. \ref{eq:e_star} is sufficient to properly account for all of the energy changes in the runs. At the later time $t\_2$ with a binary separation of $\approx 900 \, R_{sch}$ the contribution of the 2PN term to the energy of the binary starts to be visible. We can conclude that the energy discrepancy is resolved by the inclusion of the PN correction to the orbital energy relation of the binary. When the separation between the black holes exceeds $1000 \, R_{sch}$, the leading 1PN corrections are sufficient for our SMBH binary case. Below this separation the contribution of the 2PN energy correction term  also need to be taken into account to derive the binary orbital energy changes.

\section{Summary and conclusions}
\label{sec:conc}
\par In this work we have described two sets of three-body simulations that were performed using the highly accurate AR-chain regularization integration method \citep{Mikkola2006, Mikkola2008}. The goal of the simulations was to explore the interaction and study the energy exchange between a stellar particle and an SMBH binary in the final phase of the merger with the star acting as an incoming third body. More specifically, we aimed to analyze the energy balance of individual interactions and investigate the effect of Post-Newtonian terms on the binary evolution when the semi-major axis of the black hole orbit is $\approx 1300 R_{sch}$ in the case of the first set, and when the separation  is $\approx 900 R_{sch}$, in the case of the second simulation set.

\par The energy balance analysis showed that there is a significant error in the energy balance of the system when the Newtonian definition of binary orbital energy is used. The detected energy discrepancy was comparable to the other energy changes of the system and was orders of magnitude larger than the energy error of the code. The introduction of the 1PN term to the energy definition has proven sufficient to account for all energy discrepancies for separations larger than $1000 R_{sch}$.  For this reason, the orbital energy of the binary must be modified by at least the leading PN order to properly account for all energy changes in the system.  The Newtonian approximation is shown to fail even in the regime when GW emission is not the dominant effect driving the binary evolution. Alternatively, a quasi-Keplerian definition of orbital elements can be used to avoid the issue. The 1PN term (Eq. \ref{eq:epn1}), while not effectively carrying away energy from the system, does produce variations in the energy and the binary orbital parameters. The oscillations of the PN correction terms during the periodical motion of the SMBHs depend on the mass ratio and the eccentricity. Fig. \ref{fig:bind_corr} shows that the sign of the 1PN energy term is correlated with the Newtonian eccentricity change of the SMBH binary orbit.  This is likely correlated to the fact that while PN dynamics admit the conservation of angular momentum, the magnitude of the vector $\vec{r}\times\vec{v}$ is no longer conserved \citep{Will2011}, resulting in the observed changes in the eccentricity.

\begin{acknowledgements}
\par  This work was funded by a \emph{“Landesgraduiertenstipendium"} of the University of Heidelberg. It is also partly funded by the Volkswagen Foundation under the Trilateral Collaboration Scheme (Russia, Ukraine, Germany) project titled ("Accretion Processes in Galactic Nuclei") (funding for personnel and international collaboration exchanges).  The authors gratefully acknowledge the Gauss Centre for Supercomputing e.V. (www.gauss-centre.eu) for funding this project by providing computing time through the John von Neumann Institute for Computing (NIC) on the GCS Supercomputer JUWELS at Jülich Supercomputing Centre (JSC). The  authors  also  acknowledge  support  by  the  state  of Baden-W\"urttemberg through bwHPC.
PB acknowledges support by the Chinese Academy of Sciences
through the Silk Road Project at NAOC, the President’s
International Fellowship (PIFI) for Visiting Scientists
program of CAS, the National Science Foundation of China
under grant No. 11673032.

This work was supported by the Volkswagen Foundation
under the Trilateral Partnerships grants No. 90411 and 97778.

The work of PB was supported under the special program of the NRF 
of Ukraine ‘Leading and Young Scientists Research Support’ - 
"Astrophysical Relativistic Galactic Objects (ARGO): life cycle of 
active nucleus",  No. 2020.02/0346.

\end{acknowledgements}

\bibliographystyle{aa}
\bibliography{avramov_PN}

\begin{appendix} 
\section{Initial velocity condition}
\label{sec:6_vini}
\par The closest approach of the stellar particle to the binary (its pericenter distance $r_p$)  depends on its initial velocity amplitude $V_i$. Therefore, in order to make sure that our encounters experience strong interactions with the black hole binary, we need to carefully choose the initial velocity of the stellar particles.  We are interested in the particles which have closest approaches comparable to the initial semi-major axis of the binary: 
\begin{equation}
\label{eq:per_cond2}
r_p<2a_0.
\end{equation}

Since the SMBH binary is usually deeply embedded in the galaxy, we are interested in a large range of initial velocities of the stars exceeding the escape velocity of the isolated SMBH binary. We do not start with a physical distribution function, but use a simplified model instead.

 \par Since the stellar particle starts at a large distance  compared to the separation of the SMBH binary and we are only interested in a rough estimate of the pericenter distance, we replace the binary by a single object with mass $m$ at the centre of mass position. 
With the standard procedure for a two-body system, we can use energy and angular momentum conservation of the relative motion of a particle with reduced mass $\mu$ in a fixed point-mass potential of the total mass $M=m+m\_*$. Specific energy and anglular momentum are given by
\begin{align}
E\_* &= \frac{1}{2}  V\_i^2 - \frac{M}{r\_i}\\
L\_* &= | \vec{r}\_i\times \vec{V}\_i|
\end{align}
with $G=1$.
An evaluation at the initial position (index i) and at the pericenter passage (index f) leads to
\begin{align}
\label{eq:e_cons}
\frac{1}{2}  V\_i^2 - \frac{M}{D\_0} &= \frac{1}{2}  V\_f^2 - \frac{M}{r\_p} \\
\label{eq:l_cons}
V\_{i}D\_0\sin\alpha&= r\_{p}V\_{f}
\end{align}
where $\alpha$ is the initial angle between vectors $\vec{V}\_i$ and $\vec{r}\_i$ and $r\_i=D\_{0}$, $r\_f=r\_p$.
 Combining equations  \ref{eq:e_cons} and \ref{eq:l_cons} by eliminating $V_f$,  we can
 solve the quadratic equation for $r\_{p}$ and
 rewrite condition \ref{eq:per_cond2} as: 
\begin{equation}
\label{eq:rp_form}
r\_{p}=\frac{\sqrt{1+\frac{V\_{i}^2D\_{0}^2\sin^2\alpha}{M}\left( \frac{V\_{i}^2}{M}-\frac{2}{D\_{0}}\right)}-1}{\frac{V\_{i}^2}{M}-\frac{2}{D\_{0}}} \leq 2a\_{0}. 
\end{equation}
This inequality can be solved for the initial velocity amplitude, with only one unknown variable, the angle $\alpha$:

\begin{align}
\label{eq:v_cond2}
V\_{i}^2 &\leq V\_{max}^2 =\frac{4a\_{0}(D\_{0}-2a\_{0})M}{D\_{0}(D\_{0}^2\sin^2\alpha-4a\_0^2)}, \\
\rightarrow \left(\frac{V\_{i}}{V\_{esc}}\right)^2 &\leq \frac{2a\_{0}(D\_{0}-2a\_{0})}{D\_{0}^2\sin^2\alpha-4a\_0^2},
\end{align}
where we have expressed velocity in terms of the escape velocity needed for the star to be ejected from the system, $V\_{esc} = \sqrt{2M/D\_0}$. The last expression corresponds to Eq. \ref{eq:v_cond} and is useful only for angles with $|\sin\alpha|\, > 2a_0/D_0$. For smaller angles all stars come close to the SMBH binary. Since the critical angle is very small, we do not need to apply an absolute maximum for the initial velocity.


In the first step we draw the angle $\alpha$ from an isotropic distribution by defining a unit vector for the velocity. We generate a randomly oriented normalized velocity vector  $\vec{\hat{v}}$ to find the cartesian components:  

\begin{align}
v_{x}=v\_{t}\cos(\phi),\\
v_{y}=v\_{t}\sin(\phi),\\
v_{z}=R_{3},
\end{align}

where $R_{3} \in [-1,+1]$ is a randomly generated number. The quantities $v\_{t}$ and $\phi$ were calculated using
\begin{align}
\phi = 2\pi R_{4},\\
v\_{t} = \sqrt{1^2-v_{z}^{2}},
\end{align}
where $R_{4} \in [0,+1]$.

With the normalized velocity vector calculated, we can find the angle $\alpha$ using the standard scalar product: 
\begin{equation}
\cos\alpha = \frac{\vec{r}\_i}{\|\vec{r}\_i\|}\cdot \vec{\hat{v}} = \frac{x_i v_{x}+y_i v_{y}+z_i v_{z}}{D_0}.
\end{equation}

With the angle $\alpha$ calculated, we can substitute the above relation in  Eq. \ref{eq:v_cond2} in order to obtain $V\_{max}$. Then, we choose a random value $R_{5} \in [0,+1]$ to calculate $V\_i=R_{5}\,V\_{max}$ and multiply the obtained velocity amplitude with the already generated unit velocity vector $\vec{\hat{v}}$
\begin{equation}
\vec{V}\_i = \pm V\_i\vec{\hat{v}}.
\end{equation}
The sign is chosen such that the star is moving towards the SMBH binary in order to save computation time.
\end{appendix}
\end{document}